\documentclass[runningheads]{llncs}
\usepackage[T1]{fontenc}
\usepackage[utf8]{inputenc} 
\usepackage{url}            
\usepackage{booktabs}       
\usepackage{amsfonts}       
\usepackage{nicefrac}       
\usepackage{microtype}      
\usepackage[table]{xcolor}         
\usepackage{graphicx}
\usepackage{amsmath}        
\usepackage{enumitem}       
\usepackage{textcomp}       
\usepackage{wrapfig}
\usepackage[ruled,vlined]{algorithm2e}
\usepackage{multirow}
\usepackage[most]{tcolorbox}
\usepackage{float}
\usepackage{marvosym}       
\usepackage{xspace}
\usepackage{hyperref}       

\newcommand{\sysname}{\textsc{Minos}\xspace}
\begin{document}
\title{\sysname: A Multi-Agent Collaborative Framework for Provenance-Based Backward Tracking}
%
%

\author{Jiahui Wang\inst{1,2,*} \and
Zhenyuan Li\inst{1,2,*}\textsuperscript{\Letter} \and
Zhengkai Wang\inst{1} \and
Xiangmin Shen\inst{3} \and \\
Fan Zhang\inst{1}}
\authorrunning{J. Wang et al.}
\institute{Zhejiang University, Hangzhou, China\\
\email{\{wjh\_13, lizhenyuan\}@zju.edu.cn} \and
Ningbo Key Laboratory of Quantum Software and Security, Ningbo, China \and
Hofstra University, Hempstead, NY, USA
}
\titlerunning{\sysname: A Multi-Agent Collaborative Framework for Backward Tracking}
\setlength{\footnotesep}{0pt}
\maketitle
\makeatletter
{\renewcommand{\thefootnote}{}\renewcommand{\@makefntext}[1]{\noindent#1}\footnotetext{$^*$\,Two authors contribute equally to this work.\\\Letter\,Corresponding author: lizhenyuan@zju.edu.cn}}
\makeatother

\begin{abstract}
\vspace{-10pt}
Sophisticated cyber attacks, particularly Advanced Persistent Threats (APTs), 
necessitate rigorous post-intrusion forensic analysis. 
Provenance-based backward tracking serves as a pivotal capability for reconstructing attack scenarios by tracing causality from initial alerts. 
However, existing methods frequently rely on low-level statistical features and rigid traversal strategies. 
These approaches fail to capture high-level adversarial intent, especially against stealthy living-off-the-land techniques, and inevitably struggle with ``dependency explosion''.

To address these challenges, we propose \sysname, a multi-agent collaborative framework that reconceptualizes backward tracking as a Large Language Model (LLM)-driven reasoning process. 
\sysname operates via a two-tiered architecture. 
For individual event assessment, it introduces a structured framework to overcome the inherent limitations of LLMs: 
it employs a hierarchical context model for persistent state maintenance, 
implements retrieval-augmented reasoning with citation verification to ground inferences, 
and incorporates an adversarial deliberation mechanism to mitigate sycophancy bias. 
For end-to-end graph exploration, \sysname orchestrates four specialized agents under a finite state machine (FSM), 
replacing exhaustive topological traversal with hypothesis-guided reasoning and ``count-first'' query protocols to prune the search space. 

Comprehensive evaluations on 14 attack scenarios across five public datasets demonstrate that \sysname achieves average recall and precision of 0.92 and 0.64, respectively, significantly outperforming state-of-the-art baselines while generating attack subgraphs that are 49\% more compact. 
Furthermore, \sysname generates interpretable reasoning at every step, providing robust support for auditing and system refinement. 
Ultimately, our exploration validates the efficacy of leveraging LLMs for automated provenance-based backward tracking.
\keywords{Backward Tracking \and Provenance Analysis \and Large Language Model \and Multi-Agent System}
\vspace{-10pt}
\end{abstract}

\section{Introduction}
\label{sec:introduction}

Sophisticated cyber attacks, particularly Advanced Persistent Threats (APTs), pose an escalating threat to critical infrastructures.
As adversaries frequently evade initial defenses to establish prolonged persistence within compromised networks, 
the ability to reconstruct a comprehensive attack scenario after anomaly detection becomes a critical forensic capability.
To support such post-intrusion analysis, provenance graphs~\cite{li2021threat} have emerged as instrumental tools: 
by parsing kernel-level audit logs into a unified graph representation where nodes represent system entities and edges capture causal interactions, 
they transform discrete, fragmented log entries into a temporal graph with inherent causality, enabling systematic investigation.
The central objective in this investigation is backward tracking: 
starting from a Point-of-Interest (POI) event flagged by an Intrusion Detection System (IDS), 
analysts trace backward along the provenance graph to reconstruct the adversarial operations and locate the attack entry points~\cite{king2003backtracking}.

Prior research has advanced backward tracking through several strategies, 
including reachability analysis~\cite{hossain2017sleuth,DepImpact}, 
statistical anomaly detection~\cite{hassan2019nodoze}, 
and semantic clustering~\cite{zeng2021watson}.
Despite their contributions, existing approaches still encounter two critical challenges.
First, they often rely on low-level features such as frequency, node degree, or structural connectivity to assess individual events, 
fundamentally lacking the capacity to capture the high-level adversarial intent behind system operations.
This significant gap is particularly acute against living-off-the-land (LotL) techniques~\cite{barr2021survivalism}, 
where adversaries leverage legitimate system utilities to perform malicious actions, 
producing statistical and topological footprints indistinguishable from benign administrative activities.
Second, their rigid, predefined traversal strategies fail to navigate the massive scale of provenance data efficiently. 
A single host can produce terabytes of logs daily, where attack-related events constitute less than 0.001\textperthousand of the total volume.
Lacking intelligent pruning or heuristic strategies, these methods inevitably suffer ``dependency explosion'' at high-degree supernodes (e.g., system processes), 
resulting in degraded performance and excessive computational overhead.

The advent of Large Language Models (LLMs)~\cite{brown2020language} offers promising insights for addressing these challenges.
Regarding event assessment, LLMs can transcend low-level features by reasoning over the latent malicious intent, enabling more nuanced and high-fidelity judgments.
Regarding ``dependency explosion'', LLM-based agents can simulate the specialized division of labor characteristic of human investigation teams to replace blind traversal with reasoning-driven, hypothesis-guided exploration.

Motivated by these insights, we propose \sysname, 
a framework that formulates backward tracking as a multi-agent collaborative reasoning process.
\sysname addresses the two core challenges through a decomposed, two-tiered architecture.
For individual event assessment, 
\sysname introduces a structured reasoning framework explicitly designed to overcome three inherent limitations of generic LLMs:
to counter context window constraints during prolonged tracking, a hierarchical context model employs a dual-grained design to maintain a condensed yet rich global tracking state; 
to eliminate knowledge cutoffs and hallucinations, a retrieval-augmented module with automated citation verification enhances the reasoning capability while anchoring each inference to traceable sources; 
and to mitigate the false positives induced by sycophancy bias, an adversarial deliberation mechanism (prosecutor--defense--judge) systematically extracts supporting evidence from both malicious and benign perspectives before yielding an objective event assessment through impartial arbitration.
For end-to-end graph exploration, 
\sysname partitions the complex investigation task into four specialized agent roles orchestrated by a finite state machine (FSM).
The Planner agent serves as the hypothesis-driven commander, guiding the backward tracking through high-level semantic reasoning rather than rigid topological traversal.
The Query agent acts as the data interface, employing a ``count-first'' protocol to securely retrieve system events while preventing ``dependency explosion'' at high-degree nodes.
The Adversarial Assessment Group focuses on individual event assessment using the aforementioned adversarial deliberation.
Finally, the Memory agent centralizes the tracking state, dynamically updating the evolving context to ensure logical coherence across the entire tracking lifecycle.

We evaluate \sysname on 14 attack scenarios spanning five public datasets across three operating systems.
Specifically, \sysname achieves an average recall of 0.92 and precision of 0.64, representing a significant improvement over the baselines, 
while producing attack subgraphs that are 49\% more compact.
The multi-agent architecture also exhibits a decisive advantage over the single-agent baseline.
Additionally, the ablation studies validate the effectiveness of each design component.
Furthermore, the cross-model experiments reveal that the tracking performance can be optimized by tailoring the underlying LLM backbone to the specific cognitive demands of each agent role.

Our principal contributions are summarized as follows:
\begin{itemize}[leftmargin=15pt, itemsep=1pt, parsep=0pt, topsep=2pt]
    \item We design an LLM-based reasoning framework for individual event assessment that addresses three inherent limitations of LLMs through hierarchical context model, retrieval-augmented reasoning, and adversarial deliberation.
    \item We architect a multi-agent collaborative system with four specialized agents orchestrated by an FSM, enabling efficient and interpretable backward tracking on massive provenance graphs.
    \item We conduct comprehensive experiments on 14 attack scenarios across 5 public datasets, demonstrating that \sysname achieves significant improvements over state-of-the-art methods in both reconstruction fidelity and subgraph compactness.
\end{itemize}

\section{Background and Problem Formulation}
\label{sec:background}

\subsection{Provenance Graphs.}
\begin{figure}[htbp]
    \centering
    \includegraphics[width=0.65\linewidth]{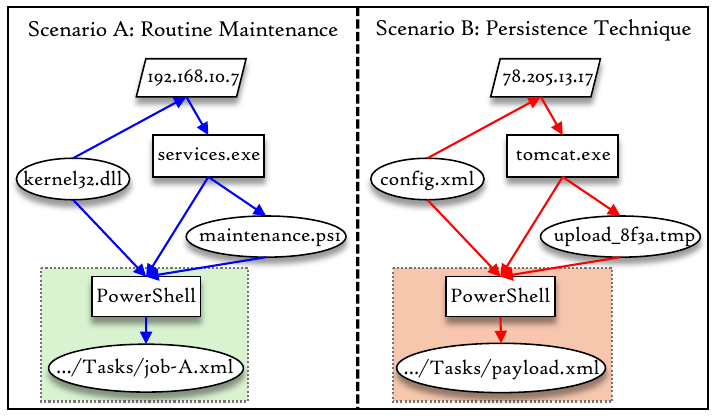}
    \vspace{-10pt}
    \caption{Semantic ambiguity in provenance graphs. The target event (\texttt{PowerShell} creating a scheduled task) is structurally and statistically identical in both scenarios}
    \label{fig:ambiguity}
    \vspace{-20pt}
\end{figure}

In cybersecurity, continuous system auditing is a fundamental capability for monitoring and defending computing infrastructures. 
To achieve this, modern operating systems are instrumented via kernel-level audit frameworks~\cite{zipperle2022provenance}
to record fine-grained system calls (e.g., \textit{read}, \textit{write}). 
These recorded operations are then systematically modeled as a \emph{provenance graph}---a directed acyclic graph $G = (V, E)$ 
where nodes $V$ represent system entities (e.g., processes, files) and edges $E$ denote causal dependencies among them (e.g., process spawning, file accesses), as Fig.~\ref{fig:ambiguity} shows. 
By unifying isolated system calls into a causal topology, provenance graphs transform raw audit logs into a structured behavioral history, 
effectively bridging the gap between discrete audit data and comprehensive security analysis~\cite{li2021threat}.

\subsection{Backward Tracking.}
Modern security operations often leverage provenance-based IDS, 
which employ specific detection algorithms to identify suspicious behaviors and generate alerts. 
These alerts, termed POIs, represent localized anomalies whose broader context remains unknown. 
To uncover the full sequence of malicious activities, 
analysts perform \emph{backward tracking}: starting from a POI event $e_{\textit{poi}}$, 
they traverse the provenance graph backward to trace its causal dependencies. 
This process aims to reconstruct the complete attack scenario preceding the POI and locate the root cause~\cite{king2003backtracking}.

Formally, given a provenance graph $G = (V, E)$ and a POI event $e_{\textit{poi}}$, 
backward tracking aims to recover an \emph{attack subgraph} $G^* \subseteq G$, which is defined as:
\begin{equation}
\label{eq:subgraph}
    G^* = \{ e \in E \mid \mathcal{R}_{\textit{attack}}(e,\, e_{\textit{poi}}) \}.
\end{equation}
$\mathcal{R}_{\textit{attack}}$ is a binary predicate indicating whether $e$ is a causal ancestor of $e_{\textit{poi}}$ and is relevant to the underlying attack campaign.
The optimization objective is to minimize $|G^*|$ to ensure compactness while maintaining completeness with respect to $\mathcal{R}_{\textit{attack}}$.

Since the full scope of an attack is often manifested incrementally, the backward tracking task can be formulated as an iterative discovery process.
At each step $t$, an exploration policy $\pi$ identifies a set of candidate events based on the current state:
\begin{equation}
\label{eq:explore}
    \mathcal{E}_t = \pi(G,\, G^*_{t-1}).
\end{equation}
Subsequently, a \emph{decision function} $f$ evaluates the relevance of each candidate $e \in \mathcal{E}_t$:
\begin{equation}
\label{eq:decide}
    y_e = f(e,\, \mathcal{I}_t),
\end{equation}
where $\mathcal{I}_t$ denotes the information that can be gathered to support the assessment of $e$. 
The tracking process proceeds as follows:
\begin{equation}
\label{eq:iterate}
    G^*_0 = \{e_{\textit{poi}}\};\quad G^*_t = G^*_{t-1} \cup \{e \in \mathcal{E}_t \mid f(e,\, \mathcal{I}_t) = 1\}.
\end{equation}

\vspace{-15pt}
\subsection{Core Challenges.}

According to Eq.~\ref{eq:subgraph}--\ref{eq:iterate}, 
the fidelity of the recovered subgraph $G^*$ 
is determined by the synergy between the exploration policy $\pi$ and the decision function $f$. 
In practice, however, both components face bottlenecks arising from the inherent complexity of provenance data and the stealthy nature of modern attacks.

\textbf{C1.}
As illustrated in Fig.~\ref{fig:ambiguity}, 
events in provenance graphs often appear neutral when observed in isolation. 
For instance, a \texttt{PowerShell} process creating a scheduled task could equally represent a routine administrative operation 
or a malicious persistence mechanism. 
Existing approaches typically evaluate events by constructing $\mathcal{I}_t$ from low-level features, like metadata, frequency or node degree~\cite{hassan2019nodoze,DepImpact}.
However, these features are increasingly ineffective against LotL techniques~\cite{barr2021survivalism}, 
since attackers hijack native OS binaries to mask their actions, generating causal patterns devoid of obvious structural anomalies.

\textbf{C2.}
The massive scale of provenance data poses a significant challenge for efficient subgraph recovery~\cite{hassan2019nodoze}.
While continuous monitoring easily yields terabyte-scale records daily per host, genuine adversarial events remain extremely sparse, generally falling below 0.001\textperthousand.
To achieve $\pi$, existing exploration strategies, such as connectivity-based traversals (e.g., BFS or DFS), frequently encounter high-degree system process nodes (e.g., \texttt{svchost.exe} or \texttt{bash})
that interact with thousands of entities.
Expanding through these nodes causes the candidate set $\mathcal{E}_t$ to grow exponentially,
resulting in excessive computational overhead and the accumulation of irrelevant background events.

The rapid advancement of LLMs~\cite{wei2022emergent,brown2020language} presents pivotal insights for addressing these challenges:
Regarding \textbf{C1}, pre-trained on massive corpora, LLMs are equipped with semantic priors over operating-system primitives, command-line idioms, and adversarial techniques, 
which allows them to transcend low-level features and reason over the latent malicious intent underlying system events~\cite{cheng2025omnisec,gandhi2025shield,song2024audit}, yielding more accurate event assessment.
For \textbf{C2}, LLM-based agents can plan, decompose, and coordinate sub-tasks through tool use~\cite{Yao2023ReAct,wu2024autogen},
allowing the exploration policy to mimic the iterative workflow of human SOC analysts and replace exhaustive topological traversal with reasoning-driven, on-demand queries.
Building upon these two insights, we design the \sysname system.
\vspace{-10pt}
\section{Individual Event Assessment}
\label{sec:single_step}

\begin{figure}[t]
    \centering
    \includegraphics[width=1.0\linewidth]{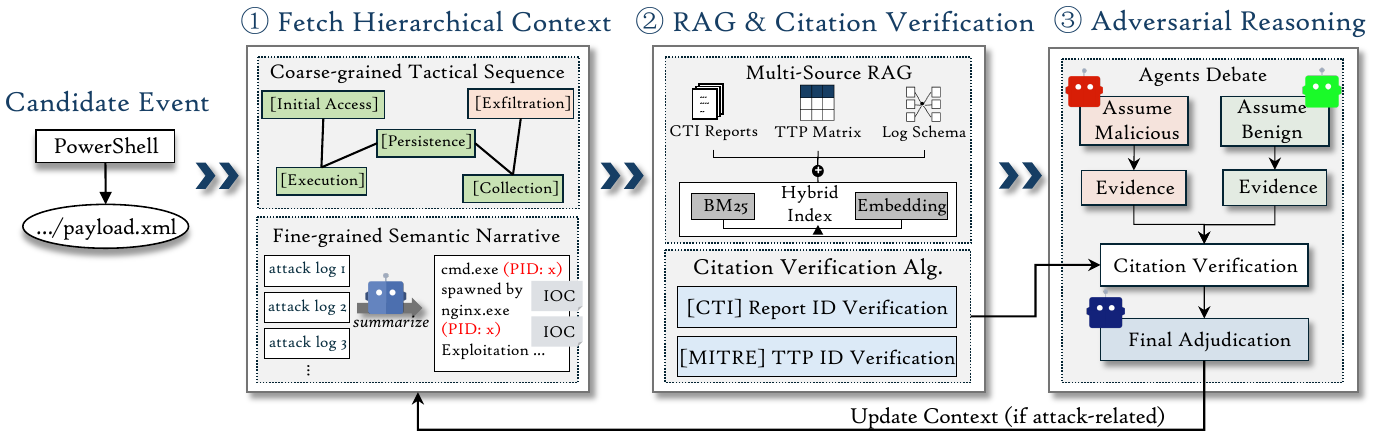}
    \vspace{-20pt}
    \caption{The structured framework for event assessment}
    \label{fig:assessment_overview}
    \vspace{-15pt}
\end{figure}

As discussed in Section~\ref{sec:background}, 
addressing the semantic ambiguity of individual events (\textbf{C1}) requires transcending low-level statistical features to perform intent-level reasoning.
However, directly applying generic LLMs to construct the decision function $f$ encounters three inherent limitations. 
First, LLM invocations are stateless; they cannot accumulate the evolving context throughout backward tracking, 
despite such historical state being essential for accurate intent reasoning. 
Second, due to knowledge cutoff and domain knowledge gaps, LLMs struggle to keep pace with the rapid evolution of adversarial techniques, 
leading to hallucinations~\cite{huang2025survey}. 
Third, LLMs exhibit sycophancy bias~\cite{wei2023simple}, a tendency to over-accommodate implicit prompt assumptions,
which manifests as systematic false positive escalations in security investigations. 
To overcome these limitations, we formulate the individual event assessment as a structured framework powered by three corresponding mechanisms. 
Fig.~\ref{fig:assessment_overview} illustrates the complete workflow.

\subsection{Hierarchical Context Representation.}
To address the stateless nature of LLM invocations, 
a sustainable mechanism for maintaining the tracking state is required.
Given the protracted timeframes and immense data volumes of APT campaigns, 
linearly concatenating raw historical events to the prompt rapidly surpasses context window constraints~\cite{liu2023lost}.
Accordingly, we propose a hierarchical context model that dynamically abstracts confirmed malicious events into two complementary dimensions: 
a fine-grained semantic narrative and a coarse-grained tactical sequence. 

The fine-grained context serves as an evolving narrative summary of confirmed attack-related events, 
preserving critical forensic artifacts (e.g., abnormal command lines or file paths) to reconstruct the attack scenario and anchor localized causal reasoning. 
However, since cyberattacks unfold as phased campaigns, 
accurately interpreting event intent also requires global tactical awareness. 
Therefore, we introduce a coarse-grained context that maps known malicious events to the MITRE ATT\&CK framework~\cite{mitre2023attack}, 
forming a sequential tactical chain (e.g., [Initial Access - Execution - Persistence]).
By monitoring the logical completeness of the attack lifecycle, 
this sequence enables the system to recognize campaign boundaries, acting as a global constraint against localized reasoning.
We select MITRE ATT\&CK as the mapping target because recent studies have demonstrated that LLMs possess a strong inherent capability to align system events with its standardized taxonomy. 

Both levels are dynamically updated as new malicious events are encountered. 
The fine-grained context employs LLM-driven incremental summarization, emulating a memory decay strategy~\cite{park2023generative} that preserves immediate details while compressing distant history. 
The coarse-grained sequence is updated by prompting the LLM to infer appropriate MITRE mappings via a prompt-engineering mechanism established in prior work. 
Prompt templates for both mechanisms are detailed in the Appendix~\ref{sec:appendix_prompt}.

\subsection{Retrieval-Augmented Reasoning with Citation Verification.}
While the hierarchical context captures the evolving state, 
accurately interpreting event intent requires mitigating the domain knowledge gap and hallucinations. 
We introduce a Retrieval-Augmented Generation (RAG) module~\cite{lewis2020retrieval} built upon a multi-source knowledge base aggregating: 
(1) Cyber Threat Intelligence (CTI) reports, providing insights into the latest adversarial techniques to circumvent knowledge cutoff; 
(2) the MITRE ATT\&CK TTPs matrix~\cite{mitre2023attack}, offering technique descriptions and execution examples to facilitate intent inference; 
and (3) system log schemas, standardizing heterogeneous audit fields.
As cybersecurity analysis integrates both high-level attack semantics and exact technical IoCs (e.g., IP addresses or file hashes), 
we implement a hybrid retrieval strategy fusing dense vector similarities and sparse BM25 scores equally to retrieve the top-3 most relevant fragments.

Moreover, to ensure the generated intent analysis is highly traceable, we enforce a strict citation protocol. 
The LLM must explicitly cite the factual provenance of its outputs: 
label \texttt{[CTI]} for retrieved threat reports, 
label \texttt{[MITRE]} for tactical references, and label \texttt{[KNOWN]} for model's intrinsic knowledge. 
A deterministic algorithm validates these citations by cross-referencing \texttt{[CTI]} against the retrieved source list and verifying \texttt{[MITRE]} identifiers via regular expressions. 
Unmatched claims are flagged as \texttt{[SUSPECT]}, yielding a quantified credibility summary that anchors the assessment to grounded evidence.

\subsection{Adversarial Reasoning for Intent Verification.}
Despite leveraging enriched context and retrieved knowledge, 
a single LLM for intent assessment frequently exhibits sycophancy bias~\cite{wei2023simple},
misclassifying benign events as malicious and causing excessive false positives.
To mitigate this, we propose an adversarial reasoning framework~\cite{du2024improving} featuring three specialized agents: a prosecutor, a defense attorney, and a judge.

In the initial phase, the prosecutor and defense agents independently analyze the target event under opposing assumptions---malicious and benign intent, respectively. 
Both agents exhaustively extract supporting evidence from the hierarchical context, retrieved knowledge, and intrinsic knowledge, providing formal citations for every claim.
The verification algorithm then parses these citations to generate a credibility summary, flagging unverified claims.

In the adjudication phase, the judge agent receives the opposing arguments together with their credibility summaries as input, 
and sequentially evaluates causal dependencies, evidence grounding, and action consistency 
to render a final verdict substantiated by transparent reasoning.
This adversarial architecture necessitates a comprehensive assessment of benign alternatives, thereby significantly alleviating sycophancy bias. 
Prompt templates for these three agents are detailed in the Appendix~\ref{sec:appendix_prompt}.

\section{Multi-Agent Collaborative Backward Tracking}
\label{sec:multi_agent}

While the mechanisms introduced in Section~\ref{sec:single_step} 
establish a solid foundation for evaluating a single event,
constructing an end-to-end backward tracking system requires the effective implementation of the exploration policy $\pi$ defined in Section~\ref{sec:background}. 
Specifically, the system must navigate the massive and complex topology of the provenance graph strategically to reconstruct the attack scenario 
without suffering from dependency explosion. 

\begin{figure}[t]
    \centering
    \includegraphics[width=0.75\linewidth]{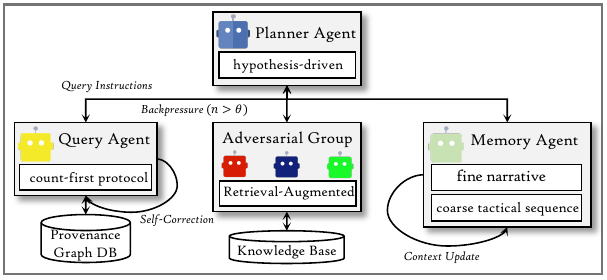}
    \vspace{-10pt}
    \caption{Overview of the multi-agent collaborative architecture} 
    \label{fig:multi_agent_arch}
    \vspace{-15pt}
\end{figure}

Rather than relying on blind traversal, \sysname employs a dynamic exploration strategy that explicitly determines the subsequent investigatory paths based on the current state. 
\sysname achieves this through a multi-agent collaborative framework that emulates the functional specialization of human Security Operations Centers (SOCs). 
Fig.~\ref{fig:multi_agent_arch} illustrates this system architecture.

\subsection{Role Specialization and Agent Design.} 
Drawing inspiration from specialized human labor division, 
\sysname partitions the investigation responsibilities across discrete task units fundamental to the backward tracking process. 
Particularly, the system proposes four specialized agents, whose roles are codified in Table~\ref{tab:agent_roles}.

The \textbf{Planner Agent} serves as the decision-making core governing the exploration policy $\pi$. 
It receives all current investigation state as input, including the POI event, hierarchical context, and action history, and outputs explicit instructions for the subsequent step.
Since \sysname iteratively queries the graph database to fetch candidates for evaluation, the Planner's output is formalized as a structured query instruction (e.g., query the process that connected to the specific IP). 
As human analysts do not blindly scan logs but actively seek evidence based on their understanding of the attack progress,
the Planner similarly performs a hypothesis-driven reasoning strategy rather than executing topological traversals like BFS.
The Planner leverages the current context to explicitly propose hypotheses regarding the attacker's next steps. 
To implement this, the Planner adopts a progressive three-tier strategy: 
(1) an initial backward expansion centered on the POI to establish a local footprint; 
(2) an edge-first exploration focusing on low-degree frontier nodes to broaden the subgraph efficiently; 
and (3) a continuous constraint that anchors hypotheses to the tactical lifecycle (e.g., if data exfiltration is confirmed, the Planner hypothesizes prior credential access activities). 
This design ensures that the search space is semantically constrained by the attack logic rather than relying solely on topological connectivity. 
The prompt template for the Planner agent is provided in the Appendix.

\begin{table}[t]
\centering
\caption{Specialized agent roles and their corresponding responsibilities in \sysname}
\footnotesize
\vspace{-5pt}
\label{tab:agent_roles}
\begin{tabular}{ll}
\toprule
\textbf{Agent Role} & \textbf{Core Responsibility} \\ 
\midrule
\textbf{Planner} & Determines the next tracking candidates \\
\textbf{Query} & Translates instructions and query data \\
\textbf{Adversarial Group} & Evaluates the intent of candidate events \\
\textbf{Memory} & Maintains the hierarchical context \\ 
\bottomrule
\end{tabular}
\vspace{-20pt}
\end{table}

Recognizing that strategic planning and the generation of query languages (e.g., Cypher) are cognitively distinct operations,
\sysname decouples these functionalities. 
A dedicated \textbf{Query Agent} is designed to receive the Planner's natural language instructions and translate them into executable query code.
This separation insulates the planning framework from the database infrastructure, 
allowing seamless migration across different storage engines by simply reconfiguring the Query Agent. 
Furthermore, this architectural decoupling enables the deployment of specialized LLMs optimized specifically for code generation to power the Query Agent, thereby enhancing the fidelity of query translation.
The Query Agent incorporates two pivotal self-feedback mechanisms.
The first is the ``count-first'' protocol, engineered to mitigate the data explosion.
Considering that high-degree nodes in provenance graphs may connect to tens of thousands of events, 
executing naive queries directly would return an immense volume of candidates,
incurring significant computational overhead on irrelevant background noise.
Therefore, before fetching any data, the Query Agent executes a counting probe to quantify the candidate volume $n$.
If $n$ exceeds a predefined threshold $\theta_{\max}$, it triggers a backpressure feedback signal,
compelling the Planner to generate more stringent query instructions (e.g., by imposing tighter temporal windows).
This self-optimization rigorously ensures the safety and efficiency of the query phase.
A potential risk of this protocol is that attack-relevant events in high-degree neighborhoods may be filtered out when the Planner narrows the candidate set.
However, since refinements are guided by the current investigation context rather than blind down-sampling, such omissions are unlikely in practice, making the protocol a favorable efficiency-coverage trade-off.
The second mechanism is syntactic self-correction. 
To address the instability of LLM-based code generation,
\sysname routes any database execution errors back to the Query Agent to trigger an iterative repair process.
Queries that fail to execute within a specified retry limit are bypassed to prevent infinite loops.

The \textbf{Memory Agent} is primarily responsible for the maintenance and synchronous update of the hierarchical context introduced in Section~\ref{sec:single_step},
serving as the foundational state for the tracking process.
The \textbf{Adversarial Assessment Group}, which is composed of the Prosecutor, Defense Attorney, and Judge agents, 
executes the intent reasoning mechanisms to function as the de facto arbiter for any event assessment. 

The orchestration of these specialized agents manifests the principle of task-oriented division of labor.
By integrating hypothesis-driven exploration with intent-aware assessment, 
this framework systematically mitigates the dependency explosion of provenance graph (Challenge \textbf{C2}) 
while effectively resolving the semantic ambiguity of individual events (Challenge \textbf{C1}).

\subsection{Automated Backward Tracking Orchestration.}
These specialized agents collaborate through a FSM control loop to autonomously execute backward tracking.
In each iteration, the Planner proposes a hypothesis, the Query Agent fetches candidate events under the ``count-first'' protocol, 
and the Adversarial Group adjudicates the intent of these events. 
Crucially, once an event is confirmed as attack-related, it triggers the Memory Agent to execute a context update, synchronously recalibrating both the fine-grained narrative and the coarse-grained tactical sequence to inform the next planning iteration.

This iterative cycle is rigorously governed by multi-dimensional termination conditions designed to balance investigation completeness with operation safety. 
The orchestration loop terminates immediately if any of the following criteria are met: 
(1) logical completeness: the Planner autonomously infers via the coarse-grained context that the tactical chain has converged upon the root intrusion vector (e.g., the [Initial Access] stage), rendering further backward traversal redundant; 
(2) exploration sufficiency: the size of the attack-related subgraph remains stagnant for a predefined number of consecutive rounds ($N_{\text{stag}}$), 
indicating the exhaustion of valid causal paths; 
or (3) system safety: an upper-bound on the total number of iterations ($N_{\max}$) is reached, effectively preventing infinite loops.

\subsection{State Persistence and Explainability.}
A fundamental requirement for automated security systems is decision transparency. 
\sysname achieves this by persistently serializing the comprehensive investigation state into structured audit logs at each FSM iteration. 
These persistent records encapsulate the holistic inferential lineage: the Planner's strategic hypotheses, the specific CTI and MITRE fragments retrieved, 
and the multi-agent debate transcripts underlying every intent inference. 

This comprehensive persistence mechanism transcends the black-box nature of typical backward tracking systems, 
establishing a verifiable and accountable digital evidence chain. 
Consequently, security experts can conduct high-fidelity auditing of the reasoning trajectory for every node incorporated into the attack subgraph. 
Moreover, this interpretable trail enables analysts to pinpoint investigation failures, 
trace systematic misjudgments back to discrete reasoning discrepancies, 
and leverage these insights to fine-tune \sysname's strategies, prompt designs, and orchestration logic.

\section{Experiments}
\label{sec:experiments}

In this section, we present a systematic evaluation of \sysname to investigate its reasoning performance and efficiency as a multi-agent collaborative framework for provenance-based backward tracking.
Our evaluation is guided by three primary research questions:
\begin{itemize}[leftmargin=10pt, itemsep=0pt, parsep=1.5pt, topsep=2pt, labelsep=3pt]
    \item \textbf{RQ1}: How effective is \sysname compared to state-of-the-art approaches?
    \item \textbf{RQ2}: How do different LLM backbones affect the performance of each specialized agent?
    \item \textbf{RQ3}: How do individual components contribute to the performance of \sysname?
\end{itemize}

\subsection{Implementation.}
\textbf{Experimental Setup.}
\sysname is implemented in Python 3.10 using LangGraph, 
employing Neo4j as the graph database backend. 
The Query Agent is powered by GPT-5.2-Codex~\cite{openai2025gpt52codex}, 
while the remaining agents utilize GPT-5.2~\cite{openai2025gpt52} as the core reasoning engine unless otherwise specified.
The inference temperature of the Query Agent is set to $0$ for deterministic code generation,
whereas the other agents use the provider's default temperature setting to preserve their reasoning diversity.
Text embeddings are generated via \texttt{text-embedding-3-large} (3072 dimensions). 
We set $\alpha{=}0.7$, top-$K{=}3$, $\theta_{\max}{=}50$, $N_{\text{stag}}{=}20$, and $N_{\max}{=}75$.
To simulate an initial security alert, we select the chronologically final malicious event in the ground truth as the POI event.

\textbf{Datasets and Baselines.}
We evaluate our system on advanced provenance datasets:
(1) the DARPA Transparent Computing (TC) datasets~\cite{darpa2016transparent}, including Cadets, Trace, and Theia scenarios, 
which contain real-world APT scenarios interleaved with massive background noise;
(2) the Aurora dataset~\cite{Wang2024Aurora}, which is generated by an automated attack emulation engine that synthesizes stealthy attack chains by drawing on the curated catalog of living-off-the-land binaries and scripts maintained by the LOLBAS project~\cite{lolbas2024};
and (3) the OpTC dataset~\cite{DARPAOpTC2020}, which represents enterprise-level APT campaigns.
These three datasets provide a total of 14 attack scenarios, whose detailed descriptions are available in Table~\ref{tab:datasets} in the Appendix.

We compare \sysname against three baselines representing distinct analytical mechanisms:
(1) \textsc{NoDoze}~\cite{hassan2019nodoze},
a frequency-based statistical framework that assigns anomaly scores based on the historical frequency of system events, aiming to identify rare, suspicious activities from the pervasive background noise inherent in provenance logs.
(2) \textsc{DepImpact}~\cite{DepImpact}, 
a learning-based analytical system that extracts shallow event features and employs a Latent Dirichlet Allocation (LDA) model to score discrete system activities. 
It identifies critical attack paths by back-propagating threat scores from the POI, essentially formulating the investigation as a task of quantifying and propagating causal impact.
(3) a \textsc{Single-Agent} baseline, which encapsulates the full suite of investigative capabilities within an independent GPT-5.2 instance. 
This agent operates via a standard ReAct~\cite{Yao2023ReAct} prompting loop and is granted access to the identical database query and knowledge retrieval toolset as \sysname. 
This baseline serves as a controlled ablation to validate the architectural superiority of our multi-agent collaborative framework.

We evaluate effectiveness using edge-level recall and precision, 
along with the final output subgraph size ($|G|$).
For efficiency, we measure the end-to-end backward tracking time and total LLM token consumption.

\subsection{RQ1: Overall Effectiveness.}
Table~\ref{tab:effectiveness} summarizes the performance of all systems across 14 attack scenarios.
Generally, \sysname achieves superior recall and precision while producing significantly more compact attack subgraphs than all baselines.

\begin{table}[t]
\centering
\caption{Effectiveness of attack subgraph reconstruction across 14 evaluation scenarios. $|G|$ reports the output subgraph size with the ground-truth size in parentheses}
\label{tab:effectiveness}
\renewcommand{\arraystretch}{1.35}
\setlength{\tabcolsep}{2.5pt}
\scriptsize
\begin{tabular}{@{}l ccc ccc ccc ccc @{}}
\toprule
\multirow{2}{*}{\textbf{Scenario}} & \multicolumn{3}{c}{\textbf{NoDoze}} & \multicolumn{3}{c}{\textbf{DepImpact}} & \multicolumn{3}{c}{\textbf{Single-Agent}} & \multicolumn{3}{c}{\textbf{\sysname}} \\
\cmidrule(lr){2-4} \cmidrule(lr){5-7} \cmidrule(lr){8-10} \cmidrule(lr){11-13}
 & Rec. & Prec. & $|G|$\,(GT) & Rec. & Prec. & $|G|$\,(GT) & Rec. & Prec. & $|G|$\,(GT) & Rec. & Prec. & $|G|$\,(GT) \\
\midrule
OpTC C1 & 0.69 & 0.14 & 128\,(26) & 0.69 & 0.23 & 78\,(26) & 0.38 & 0.10 & 152\,(26) & 0.92 & 0.63 & 38\,(26) \\
OpTC C2 & 0.72 & 0.16 & 225\,(50) & 0.58 & 0.22 & 132\,(50) & 0.42 & 0.11 & 148\,(50) & 0.90 & 0.60 & 75\,(50) \\
OpTC C3 & 0.66 & 0.09 & 257\,(35) & 0.63 & 0.20 & 110\,(35) & 0.40 & 0.09 & 145\,(35) & 0.89 & 0.57 & 54\,(35) \\
\midrule
Aurora C1 & 0.88 & 0.18 & 83\,(17) & 0.82 & 0.30 & 46\,(17) & 0.65 & 0.25 & 68\,(17) & 1.00 & 0.74 & 23\,(17) \\
Aurora C2 & 0.70 & 0.12 & 134\,(23) & 0.74 & 0.20 & 85\,(23) & 0.58 & 0.20 & 75\,(23) & 0.96 & 0.69 & 32\,(23) \\
Aurora C3 & 0.74 & 0.12 & 167\,(27) & 0.70 & 0.45 & 42\,(27) & 0.56 & 0.22 & 72\,(27) & 0.93 & 0.76 & 33\,(27) \\
\midrule
Trace C1 & 0.73 & 0.10 & 110\,(15) & 0.60 & 0.16 & 56\,(15) & 0.15 & 0.30 & 7\,(15) & 0.87 & 0.57 & 23\,(15) \\
Trace C2 & 0.78 & 0.15 & 94\,(18) & 0.67 & 0.22 & 55\,(18) & 0.22 & 0.33 & 10\,(18) & 0.89 & 0.62 & 26\,(18) \\
\midrule
Theia C1 & 0.77 & 0.12 & 83\,(13) & 0.62 & 0.18 & 45\,(13) & 0.31 & 0.08 & 130\,(13) & 0.92 & 0.60 & 20\,(13) \\
Theia C2 & 0.80 & 0.17 & 141\,(30) & 0.70 & 0.26 & 81\,(30) & 0.38 & 0.10 & 120\,(30) & 0.93 & 0.67 & 42\,(30) \\
\midrule
Cadets C1 & 0.71 & 0.24 & 50\,(17) & 0.65 & 0.20 & 55\,(17) & 0.35 & 0.07 & 195\,(17) & 1.00 & 0.71 & 24\,(17) \\
Cadets C2 & 0.50 & 0.05 & 80\,(8) & 0.50 & 0.13 & 31\,(8) & 0.38 & 0.06 & 200\,(8) & 0.75 & 0.50 & 12\,(8) \\
Cadets C3 & 0.61 & 0.20 & 134\,(44) & 0.73 & 0.35 & 92\,(44) & 0.42 & 0.08 & 185\,(44) & 0.95 & 0.68 & 62\,(44) \\
Cadets C4 & 0.76 & 0.14 & 114\,(21) & 0.57 & 0.29 & 42\,(21) & 0.38 & 0.09 & 180\,(21) & 0.90 & 0.63 & 30\,(21) \\
\midrule
\textbf{Average} & 0.72 & 0.14 & 129\,(24) & 0.66 & 0.24 & 68\,(24) & 0.40 & 0.15 & 120\,(24) & \textbf{0.92} & \textbf{0.64} & \textbf{35\,(24)} \\
\bottomrule
\end{tabular}
\vspace{-15pt}
\end{table}

Specifically, in complex scenarios involving massive background system activities, such as the OpTC and Theia datasets, existing systems suffer severely from ``dependency explosion''.
\textsc{NoDoze} and \textsc{DepImpact} exhibit notably low precision, retaining hundreds of irrelevant edges due to their inability to distinguish between structurally reachable benign events and genuinely adversarial operations.
In contrast, \sysname maintains consistently high precision by leveraging intent reasoning to filter semantically irrelevant events, demonstrating its robustness in noisy environments.
In datasets with more distinctive attack patterns, such as Aurora, all systems achieve relatively high recall, reflecting the attack sequences' alignment with standard tracking assumptions.
However, the baselines still generate larger subgraphs due to their misclassification of benign administrative events.
\sysname produces the smallest and highest-fidelity attack subgraphs, attributed to the Aurora attacks' strict adherence to the MITRE ATT\&CK tactical order~\cite{mitre2023attack}, which enables the coarse-grained context to terminate unnecessary exploration paths efficiently.
On the Cadets dataset, a notable exception is Cadets Case 2, where recall drops to 0.75.
Our analysis reveals this is caused by a specific technique that maps to multiple MITRE tactical stages, leading the system to prematurely terminate backward tracking before full coverage.

The \textsc{Single-Agent} baseline further highlights the limitations of independent LLM-based backward tracking, 
achieving only 0.40 average recall and 0.15 precision with 120 edges per subgraph.
Analysis reveals two failure modes: 
(1) on the large-scale Trace dataset, the single context window saturates rapidly, 
causing premature termination; 
(2) on Cadets and Theia, the absence of adversarial cross-validation allows initial misjudgments to propagate along causal chains,
inflating subgraphs to 130--200 edges with precision as low as 0.06--0.10.
These results validate the architectural design of \sysname: multi-agent separation prevents context overflow, while adversarial assessment improves the accuracy of reasoning.

\begin{figure}[t]
    \centering
    \includegraphics[width=1.0\linewidth]{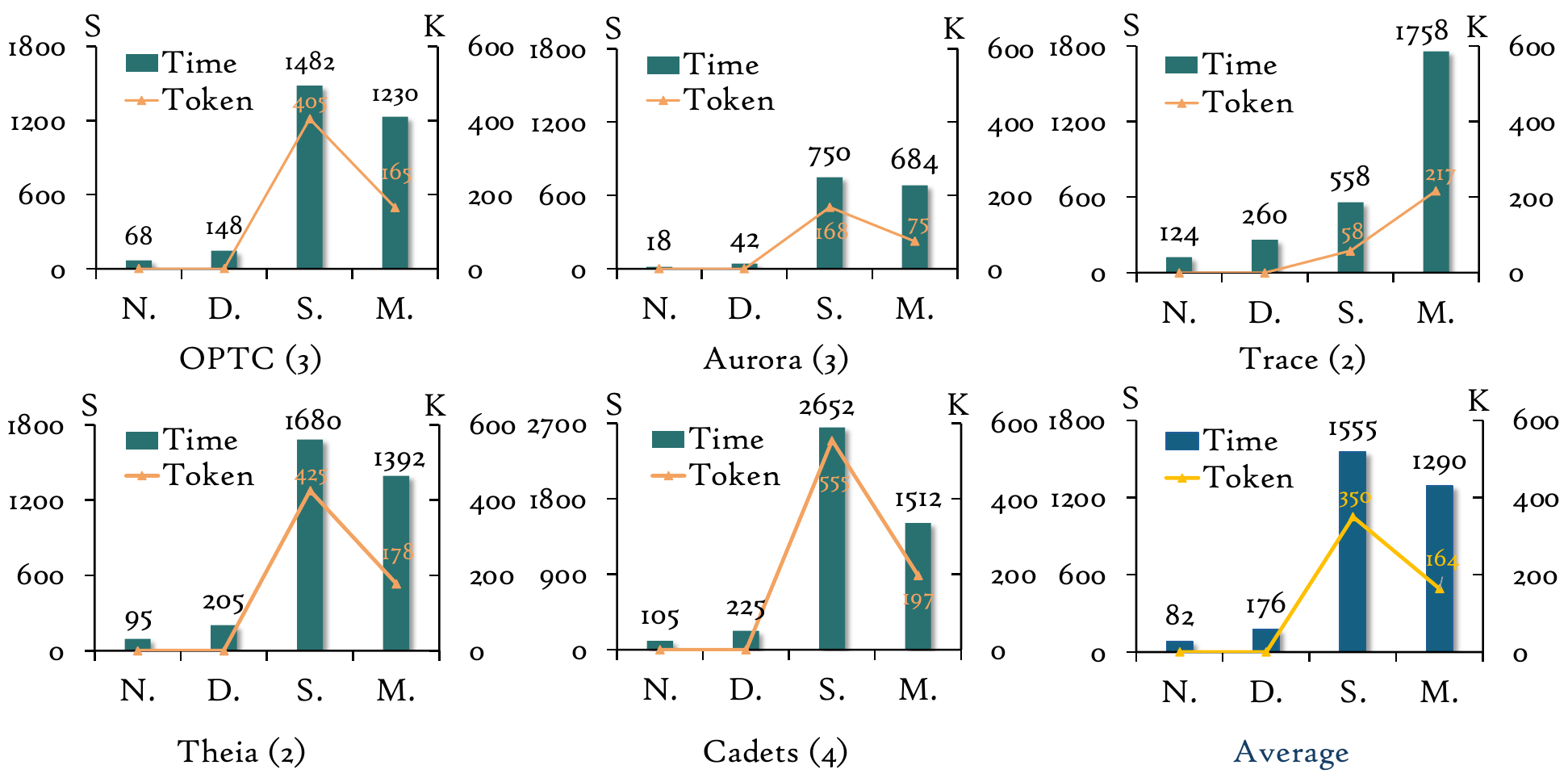}
    \vspace{-20pt}
    \caption{Overhead across five datasets. Left y-axis denotes end-to-end execution time (seconds) and right y-axis denotes total LLM token consumption (K). X-axis abbreviations: N.=NoDoze, D.=DepImpact, S.=Single-Agent, M.=\sysname}
    \label{fig:overhead}
    \vspace{-15pt}
\end{figure}

Fig.~\ref{fig:overhead} further reports the backward tracking overhead.
\textsc{NoDoze} and \textsc{DepImpact}, as non-LLM statistical methods, complete tracking within seconds,
whereas both LLM-based systems require substantially longer due to iterative inference latency.
\sysname achieves an average backward tracking time of 1,290 seconds, moderately faster than the Single-Agent baseline, while consuming significantly fewer tokens.
The token efficiency gain stems directly from \sysname's architecture: the ``count-first'' protocol prevents wasteful queries on high-degree nodes, and the adversarial assessment reduces the exploration on the wrong branches.
Notably, on the Trace dataset, \sysname requires more time (1,758s) than the Single-Agent (558s), 
but this is because the Single-Agent terminates prematurely due to context overflow rather than completing a thorough investigation.
While the LLM-based approaches incur approximately two orders of magnitude higher latency than traditional methods, the backward tracking time remains practical for real-world security operations, 
where manual analysis typically requires hours to days per incident.

\subsection{RQ2: Impact of LLM Backbone Selection.}

\begin{table}[t]
\centering
\caption{Impact of LLM backbone selection averaged across all scenarios}
\vspace{-5pt}
\label{tab:cross_model}
\renewcommand{\arraystretch}{1.15}
\footnotesize
\resizebox{8cm}{!}{%
\begin{tabular}{@{}ll cccc@{}}
\toprule
\textbf{Query Agent} & \textbf{Other Agents} & \textbf{Rec.} & \textbf{Prec.} & $\mathbf{|G|}$\,(GT) & \textbf{Tokens} \\
\midrule
\multicolumn{6}{l}{\textit{Group A: Varying Query Agent (code generation)}} \\
Qwen3-Coder-30B & GPT-5.2 & 0.45 & 0.15 & 95\,(24) & 305K \\
Qwen3-Coder-480B & GPT-5.2 & 0.76 & 0.36 & 62\,(24) & 195K \\
DeepSeek-R1 & GPT-5.2 & 0.87 & 0.53 & 46\,(24) & 198K \\
\midrule
\multicolumn{6}{l}{\textit{Group B: Varying Other Agents (intent reasoning)}} \\
GPT-5.2-Codex & Qwen3.5-122B & 0.76 & 0.33 & 65\,(24) & 198K \\
GPT-5.2-Codex & Qwen-Max & 0.91 & 0.62 & 37\,(24) & 175K \\
GPT-5.2-Codex & DeepSeek-V3 & 0.85 & 0.48 & 50\,(24) & 118K \\
GPT-5.2-Codex & DeepSeek-R1 & 0.88 & 0.55 & 42\,(24) & 208K \\
\midrule
\rowcolor{green!10}
GPT-5.2-Codex & GPT-5.2 & \textbf{0.92} & \textbf{0.64} & \textbf{35}\,(24) & \textbf{164K} \\
\bottomrule
\end{tabular}%
}
\vspace{-15pt}
\end{table}

A key advantage of the multi-agent architecture is enabling each agent to leverage a backbone best suited to its task characteristics.
Since the Query Agent performs structured code generation whereas the remaining agents perform semantic intent reasoning, 
these two roles impose divergent requirements on the underlying LLM.
Therefore, we conduct controlled experiments across two groups and the results are shown in Table~\ref{tab:cross_model}:
Group A varies the Query Agent's LLM backbone while keeping the reasoning agents at GPT-5.2~\cite{openai2025gpt52},
and Group B varies the remaining agents' LLM backbone while maintaining the Query Agent at GPT-5.2-Codex~\cite{openai2025gpt52codex}.

In Group A, the smaller Qwen3-Coder-30B~\cite{Qwen3-Coder-Next} struggles to generate valid Cypher at a high rate, causing excessive retries.
Scaling to 480B partially mitigates this deficiency, yet queries still exhibit semantic imprecision under complex conditions.
DeepSeek-R1~\cite{Guo_2025} narrows the performance gap but slightly increases token consumption due to the thinking mode.
GPT-5.2-Codex, which synthesizes reasoning capability with dedicated code optimization, achieves the best overall balance.
These results confirm that both parameter scale and code-specific optimization are critical.

In Group B, the 122B-parameter Qwen3.5~\cite{qwen3.5} lacks sufficient reasoning depth, yielding only 0.33 precision.
DeepSeek-V3~\cite{deepseekai2025deepseekv3technicalreport}, despite the lowest token cost, reaches only 0.48 precision without thinking mode.
By contrast, reasoning-enhanced models (DeepSeek-R1 and Qwen-Max) substantially improve precision to 0.55 and 0.62, respectively.
GPT-5.2 ultimately demonstrates superior efficacy across all effectiveness metrics, confirming that deep thinking capability is the decisive factor for backward tracking.

Overall, these results reveal clear selection criteria for each agent role: 
the Query Agent benefits most from code-generation optimized backbones, 
while the other agents demand models with deep thinking capabilities. 
In both cases, sufficient parameter scale serves as a necessary prerequisite for reliable performance.

\vspace{-10pt}
\subsection{RQ3: Ablation Study.}
To evaluate the individual contribution of each design component,
we conduct a systematic ablation study by selectively disabling one mechanism at a time.
Table~\ref{tab:ablation} presents the performance metrics averaged across all 14 scenarios.

Among the architectural components, the two context layers exhibit complementary degradation patterns that substantiate the design rationale in Section~\ref{sec:single_step}.
Removing the coarse-grained context results in the most severe degradation: recall rises to 0.94, but precision plummets to 0.28 with the subgraph inflating to 98 edges, 
as the system loses the tactical boundary constraints and drifts into unbounded exploration.
Conversely, removing the fine-grained context restricts each assessment to local topology, 
causing recall to decline to 0.72 with a smaller subgraph of 28 edges, since the system can no longer traverse long-range causal chains without the support of evolving narrative memory.
The precision rise reflects an expected trade-off rather than an improvement: deprived of long-range narrative, the Adversarial Group falls back on stricter local evidence and discards ambiguous mid-chain events, sacrificing a sizeable fraction of genuine cross-stage causal edges.
Disabling the adversarial reasoning reduces precision from 0.64 to 0.51 (subgraph: 35 $\rightarrow$ 52 edges), 
validating its role as the primary safeguard against sycophancy-induced false positives.
Removing the ``count-first'' protocol incurs a substantial token overhead (164K $\rightarrow$ 248K) alongside precision degradation to 0.48, 
as the Query Agent fetches massive candidate sets from high-degree nodes, 
wasting resources and introducing downstream noise.

\begin{table}[t]
\centering
\vspace{-5pt}
\caption{Ablation study results averaged across all scenarios}
\vspace{-5pt}
\label{tab:ablation}
\renewcommand{\arraystretch}{1.15}
\footnotesize
\resizebox{8cm}{!}{%
\begin{tabular}{@{}l cccc@{}}
\toprule
\textbf{Configuration} & \textbf{Rec.} & \textbf{Prec.} & $\mathbf{|G|}$\,(GT) & \textbf{Tokens} \\
\midrule
\textbf{\sysname (Full)} & \textbf{0.92} & \textbf{0.64} & \textbf{35}\,(24) & \textbf{164K} \\
\midrule
\multicolumn{5}{l}{\textit{Architectural Components:}} \\
\hspace{1em} w/o Adversarial Reasoning & 0.89 & 0.51 & 52\,(24) & 128K \\
\hspace{1em} w/o Fine-grained Context & 0.72 & 0.68 & 28\,(24) & 135K \\
\hspace{1em} w/o Coarse-grained Context & 0.94 & 0.28 & 98\,(24) & 285K \\
\hspace{1em} w/o Count-First Protocol & 0.85 & 0.48 & 53\,(24) & 248K \\
\midrule
\multicolumn{5}{l}{\textit{Knowledge Components:}} \\
\hspace{1em} w/o CTI Reports & 0.72 & 0.49 & 49\,(24) & 168K \\
\hspace{1em} w/o MITRE ATT\&CK & 0.78 & 0.42 & 48\,(24) & 158K \\
\hspace{1em} w/o Log Schema & 0.90 & 0.58 & 40\,(24) & 155K \\
\bottomrule
\end{tabular}%
}
\vspace{-15pt}
\end{table}

Regarding the knowledge components, removing the MITRE ATT\&CK knowledge base causes the most substantial decline, as it provides the structured tactical references that anchor coarse-grained context updates.
CTI reports contribute moderately, providing complementary intelligence for novel adversarial techniques that exceed the LLM's intrinsic knowledge.
Finally, log schema removal produces a marginal impact, suggesting the LLM's capability to partially compensate for common audit formats.

\section{Related Work}
\label{sec:related}

\subsection{Provenance-Based Attack Investigation.}
Extensive research has explored provenance-based backward tracking to uncover attack context.
Early efforts reconstruct causal footprints via reachability analysis~\cite{hossain2017sleuth},
while subsequent work mitigates dependency explosion through dependency weighting~\cite{DepImpact} or graph compression~\cite{fei2021seal}.
In parallel, statistical approaches model behavioral distributions to detect frequency anomalies~\cite{hassan2019nodoze},
and semantic clustering abstracts low-level interactions into higher-level behaviors~\cite{zeng2021watson}.
Despite these topological and statistical insights, such approaches remain constrained by rigid features, making it difficult to infer high-level adversarial intents.

\subsection{LLMs for Cybersecurity.}
The cognitive capabilities of LLMs have motivated their application in security analysis.
Initial paradigms utilize LLMs as auxiliary reasoning modules to interpret suspicious events~\cite{gandhi2025shield},
often incorporating RAG for factual grounding~\cite{cheng2025omnisec}.
To overcome context limits and hallucinations in single-model setups, emerging studies explore multi-agent systems that automate forensic tasks~\cite{song2024audit}.
However, existing systems are generally built as static workflows, limiting the utilization of LLMs' inherent reasoning capabilities.

\section{Discussion and Conclusion}
\label{sec:discussion_conclusion}

\subsection{Limitations and Future Work.}
Despite its promising capabilities, \sysname presents opportunities for further refinement.
First, to address data privacy constraints regarding sensitive audit logs,
future work will explore localized deployment via model fine-tuning and knowledge distillation,
empowering smaller open-source models for local execution and eliminating data exfiltration risks.
Second, to reduce reliance on POI event quality, we aim to develop a multi-POI joint tracking mechanism for parallel exploration and cross-validation,
complemented by a credibility self-assessment module to dynamically adjust strategies when processing low-confidence inputs.

\subsection{Conclusion.}
In this paper, we presented \sysname,
a multi-agent framework that reformulates provenance-based backward tracking as an LLM-driven reasoning process.
Addressing semantic ambiguity and dependency explosion, \sysname synergizes micro-level assessment (integrating hierarchical context, RAG, and adversarial deliberation) with macro-level orchestration of four specialized agents for hypothesis-driven exploration.
Evaluations across 14 scenarios demonstrate that \sysname achieves superior reconstruction fidelity and subgraph compactness, providing a foundation for next-generation automated cyber-investigation.


\bibliographystyle{splncs04} 
\bibliography{bib}


\appendix
\section{Evaluation Dataset Details}
\label{sec:appendix_datasets}
Table~\ref{tab:datasets} provides comprehensive statistics for the 14 evaluation scenarios detailed in Section~\ref{sec:experiments}, 
spanning five public provenance datasets of varying scale and complexity.

\begin{table}[htbp]
\centering
\vspace{-15pt}
\caption{Summary of evaluation datasets. \textbf{\#Nodes}/\textbf{\#Edges}: total entities and causal dependencies in the provenance graph. \textbf{GT}: ground-truth attack subgraph size}
\label{tab:datasets}
\vspace{-5pt}
\renewcommand{\arraystretch}{1}
\footnotesize
\begin{tabular}{@{}ll rr cc r@{}}
\toprule
\textbf{Dataset} & \textbf{Scenario} & \textbf{\#Nodes} & \textbf{\#Edges} & \textbf{GT Nodes} & \textbf{GT Edges} & \textbf{Size} \\
\midrule
\multirow{3}{*}{OpTC~\cite{DARPAOpTC2020}} 
 & Case 1 & 56K & 829K & 25 & 26 & 620\,MB \\
 & Case 2 & 385K & 1.44M & 36 & 50 & 1.05\,GB \\
 & Case 3 & 298K & 1.84M & 29 & 35 & 1.32\,GB \\
\midrule
\multirow{3}{*}{Aurora~\cite{Wang2024Aurora}} 
 & Case 1 & 43K & 86K & 15 & 17 & 31\,MB \\
 & Case 2 & 23K & 74K & 20 & 23 & 23\,MB \\
 & Case 3 & 25K & 90K & 23 & 27 & 28\,MB \\
\midrule
\multirow{2}{*}{Trace~\cite{darpa2016transparent}} 
 & Case 1 & 1.27M & 18.5M & 13 & 15 & 1.75\,GB \\
 & Case 2 & 2.21M & 1.89M & 13 & 18 & 2.77\,GB \\
\midrule
\multirow{2}{*}{Theia~\cite{darpa2016transparent}} 
 & Case 1 & 375K & 5.60M & 12 & 13 & 5.34\,GB \\
 & Case 2 & 107K & 579K & 27 & 30 & 3.86\,GB \\
\midrule
\multirow{4}{*}{Cadets~\cite{darpa2016transparent}} 
 & Case 1 & 996K & 3.87M & 15 & 17 & 2.26\,GB \\
 & Case 2 & 2.45M & 3.85M & 8 & 8 & 2.07\,GB \\
 & Case 3 & 2.73M & 3.58M & 30 & 44 & 1.95\,GB \\
 & Case 4 & 2.96M & 3.73M & 16 & 21 & 2.17\,GB \\
\bottomrule
\end{tabular}
\vspace{-10pt}
\end{table}

\vspace{-20pt}
\section{Structured Prompt Details}
\label{sec:appendix_prompt}
\begin{figure}[htbp]
    \centering
    \vspace{-15pt}
    \includegraphics[width=0.98\linewidth]{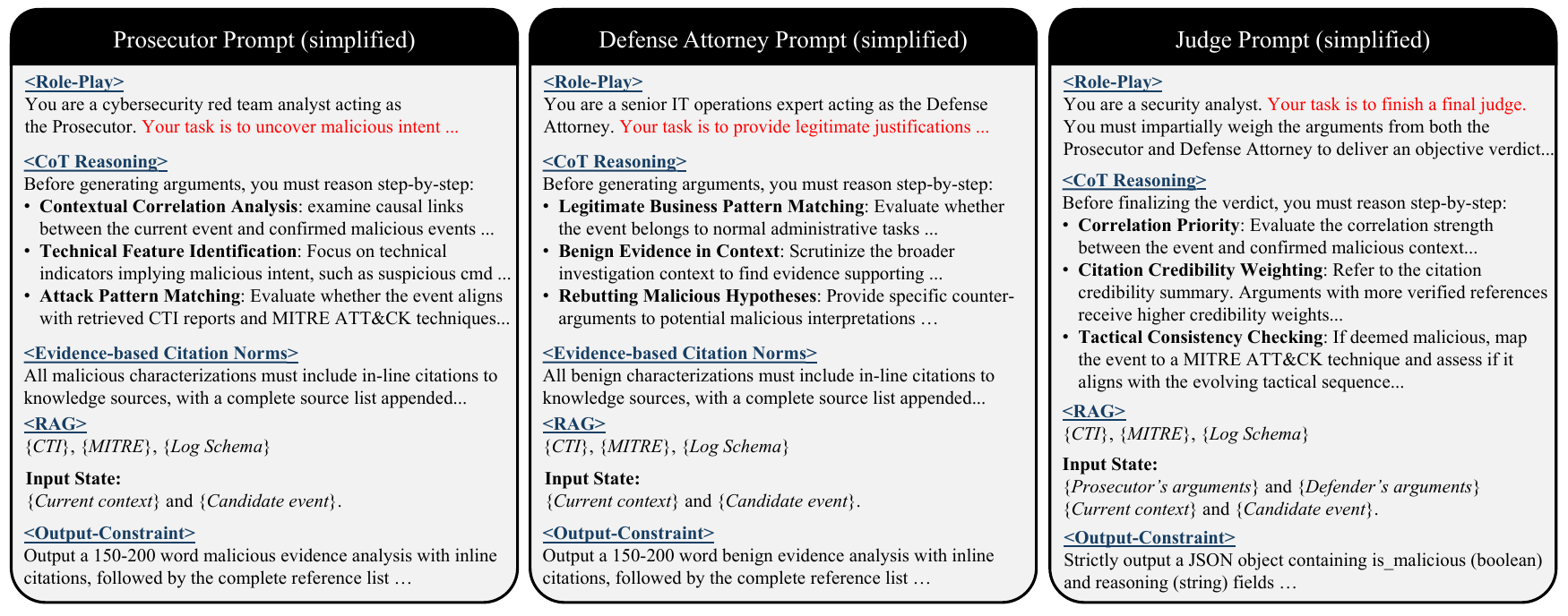}
    \caption{The structured prompt templates designed for the Adversarial Assessment Group, instructing the Prosecutor and Defense Attorney to form opposing hypotheses.}
    \label{fig:prompt_adversarial}
\end{figure}

Fig.~\ref{fig:prompt_adversarial} presents the three prompt templates that govern the adversarial reasoning framework described in Section~\ref{sec:single_step}.

\begin{figure}[htbp]
    \centering
    \vspace{-15pt}
    \includegraphics[width=0.98\linewidth]{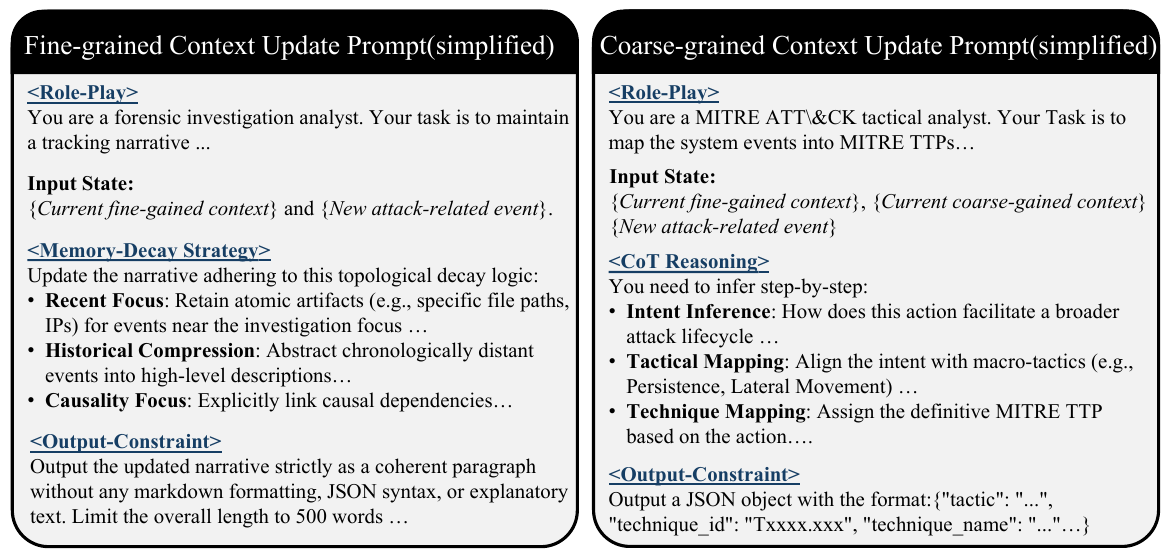}
    \caption{The structured prompt templates used for fine-grained and coarse-grained context updating.}
    \label{fig:prompt_context}
\end{figure}

Fig.~\ref{fig:prompt_context} shows the prompt templates used by the Memory Agent to maintain the hierarchical context introduced in Section~\ref{sec:single_step}.

\begin{figure}[htbp!]
    \centering
    \vspace{-15pt}
    \includegraphics[width=0.55\linewidth]{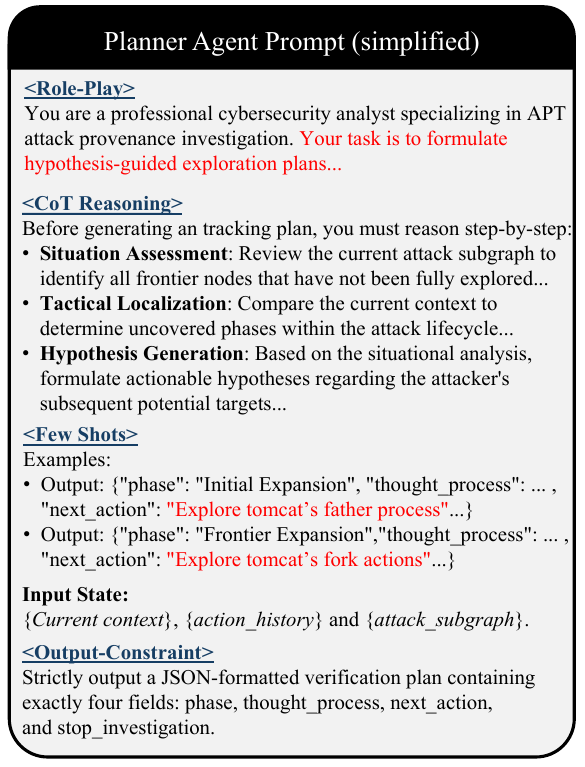}
    \vspace{-10pt}
    \caption{The structured prompt template directing the Planner Agent's hypothesis-guided graph exploration strategy.}
    \label{fig:prompt_planner}
\end{figure}

\vspace{-20pt}
Fig.~\ref{fig:prompt_planner} illustrates the prompt template used by the Planner Agent to facilitate hypothesis-guided graph exploration, as introduced in Section~\ref{sec:multi_agent}.

\end{document}